\documentclass[aps,preprint,showpacs] {revtex4}
\usepackage{graphicx}   % for figures
\usepackage{epstopdf}
\usepackage{amsmath}

\begin{document}

\title{De Vries Behavior in Smectics near a Biaxiality Induced Smectic A - Smectic C Tricritical Point}

\author{Karl Saunders}

\affiliation{Department of Physics, California Polytechnic State
University, San Luis Obispo, CA 93407, USA}

\email{ksaunder@calpoly.edu}

\date{\today}

% Last modified 2/25/08

\begin{abstract}

We show that a generalized Landau theory for the smectic $A$ and $C$ phases exhibits a biaxiality induced $AC$ tricritical point. Proximity to this tricritical point depends on the degree of orientational order in the system; for sufficiently large orientational order the $AC$ transition is 3D $XY$-like, while for sufficiently small orientational order, it is either tricritical or 1st order. We investigate each of the three types of $AC$ transitions near tricriticality and show that for each type of transition, small orientational order implies de Vries behavior in the layer spacing, an unusually small layer contraction. This result is consistent with, and can be understood in terms of, the ``diffuse cone" model of de Vries. Additionally, we show that birefringence grows upon entry to the $C$ phase. For a continuous transition, this growth is more rapid the closer the transition is to tricriticality. Our model also predicts the possibility of a nonmontonic temperature dependence of birefringence.

\end{abstract}


\maketitle

\section{Introduction}
\label{Introduction}

Since its discovery in the 1970's \cite{AC discovery}, the nature of the smectic $A$- smectic $C$ transition has been a topic of great interest. Early work showed that many systems exhibit a continuous $AC$ transition which could be described by a mean field model near tricriticality \cite{Huang&Viner}. A tricritical point, with associated neighboring 2nd order and weakly 1st order transitions was later found \cite{HuangTP, ShashidharTP}. The origin of an $AC$ tricritical point has been of significant interest, with two main mechanisms having been proposed. The first is the coupling of the tilt to biaxiality, which in chiral systems is related to the size of spontaneous polarization \cite{HuangTP, ShashidharTP}. The second is the width of the $A$ phase \cite{Huang&Lien}. Another mechanism, involving a coupling between tilt and smectic elasticity has also been proposed \cite{Benguigui}, but this seems less likely. Until now, a comprehensive theory that addresses the effect of biaxiality on the nature of the $AC$ transition has not been produced.

More recently, much attention has been given to de Vries materials, which exhibit an $AC$ transition with an unusually small change in layer spacing and a significant increase in birefringence (associated with an increase in orientational order) upon entry to the $C$ phase \cite{de Vries review}. Some de Vries materials exhibit another unusual feature, namely a birefringence that varies nonmonotonically with temperature \cite{Lagerwall, YuriPrivate}; in particular, the birefringence decreases as the $AC$ transition is approached from within the $A$ phase. De Vries materials generally seem to have unusually small orientational order and follow the phase sequence isotropic ($I$) - $A$ - $C$. In several de Vries materials, the $AC$ transition seems to occur close to tricriticality \cite{HuangDV, Hayashi1} .

Separate theoretical models \cite{Saunders, Gorkunov} have been developed, each of which predicts the possibility of a continuous $AC$ transition with the two main signatures of de Vries behavior: small layer contraction and increase in birefringence upon entry to the $C$ phase. There are differences between the assumptions used in the models, the most significant of which is the treatment of the temperature dependence of the layering order parameter; the model of Gorkunov {\it et al} \cite{Gorkunov} does not take this into account while that of Saunders {\it et al} does \cite{Saunders}. Given the absence of a nematic phase in de Vries materials, incorporating the temperature variation of the layering order parameter is of crucial importance in the modeling of de Vries materials. It seems most likely that the $IA$ transition in de Vries materials is primarily driven by the development of layering order, with orientational order being secondarily induced by the layering order. This is consistent with the general observation \cite{de Vries review} that de Vries materials have unusually strong layering order and unusually weak orientational order. Additionally, only by including temperature dependent layering, does one predict \cite{Saunders} the unusual, yet experimentally observed \cite{Lagerwall, YuriPrivate}, possibility of a nonmonotonic temperature dependence of birefringence.

Neither model considers the effect of biaxiality on the $AC$ transition. The model of Gorkunov {\it et al} investigates the possibility of an $AC$ transition that has signatures of tricriticality, but does not predict a tricritical point or the possibility of a 1st order $AC$ transition.
\begin{figure}
\includegraphics[scale=1]{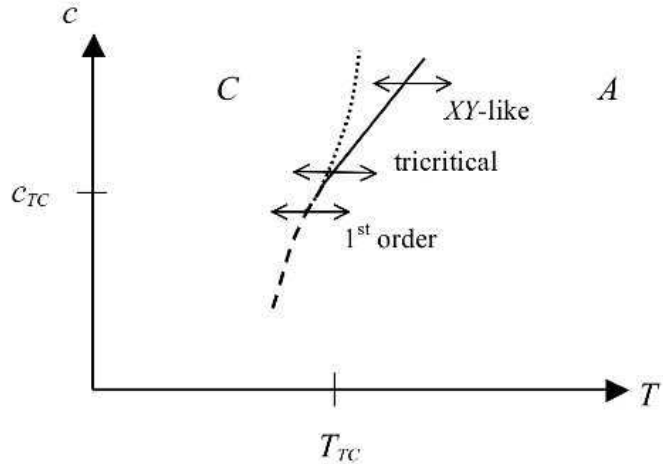}
\caption{Phase diagram in temperature ($T$) - concentration ($c$) space. For materials with excluded volume interactions, increasing the concentration would lead to an increase in the orientational order. The solid line represents the continuous $AC$ boundary while the dashed line represents the 1st order $AC$ boundary. These two boundaries meet at the tricrtical point: ($T_{_{TC}}, c_{_{TC}}$). The dotted line indicates the region in which the behavior in the $C$ phase crosses over from $XY$-like to tricritical. The region in which the behavior is $XY$-like shrinks to zero as the tricritical point is approached. Also shown as double ended arrows, are the three distinct classes of transitions (at fixed concentration): $XY$-like, tricritical and 1st order. }
\label{PhaseDiagram_T_conc}
\end{figure}

In this article, we present and analyze a new generalized nonchiral Landau theory, based on that developed in Ref. \cite{Saunders}, which includes orientational, layering, tilt and biaxial order parameters. The model naturally produces a coupling between tilt and biaxiality and we show that this coupling leads to an $AC$  tricritical point. We show that the effect of biaxiality is stronger in systems with small orientational order, $M_0$, so that a tricritical point and associated neighboring 1st order transition can be accessed by systems with sufficiently small orientational order, $M_0\leq M_{TC}$. Here $M_{TC}$ is the value of the orientational order at which the system exhibits a tricritical $AC$ transition. This means that the two mechanisms that have been proposed as leading to tricriticality, the coupling of tilt to biaxiality and the width of the $A$ phase, may in fact be two sides of the same coin. Systems with a narrow $A$ phase, which are thus close to the $I$ phase, will have small orientational order, which according to our model, leads to an enhanced effect of the biaxiality on the nature of the $AC$ transition. For materials with excluded volume interactions, a decrease in orientational order could be achieved by decreasing concentration.
\begin{figure}
\includegraphics[scale=1]{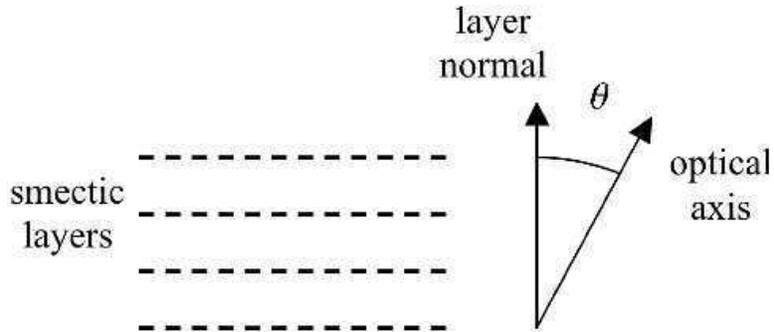}
\caption{A schematic showing the layer normal and optical axis. The layers are shown as dashed lines. The transition from the $A$ to $C$ phase occurs via a tilting, by angle $\theta$, of the optical axis away from the layer normal.}
\label{Theta_fig}
\end{figure}

\begin{figure}
\includegraphics[scale=1]{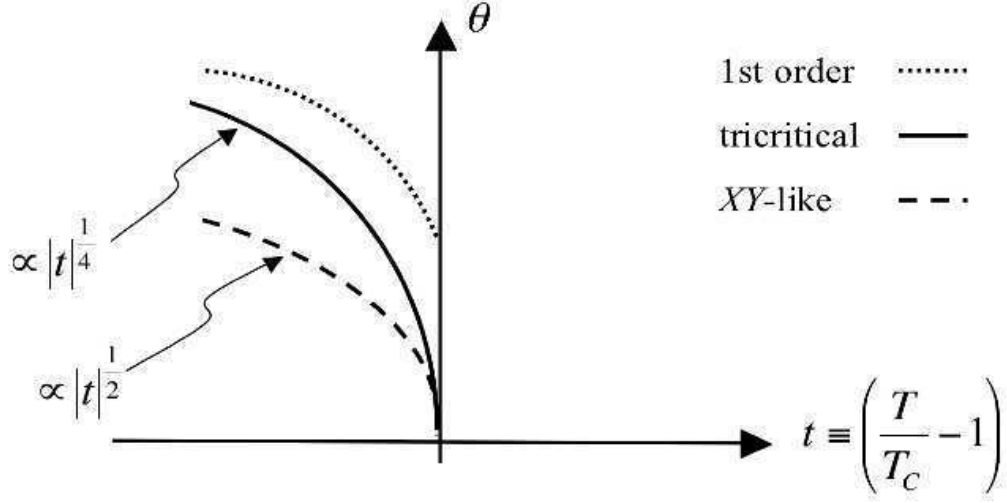}
\caption{The tilt angle $\theta$ as a function of reduced temperature $t \equiv \left(1-\frac{T}{T_C}\right)$ near the $AC$ transition temperature $T_C$, i.e., for $t \ll1$. Upon entry to the $C$ phase the growth of the tilt angle scales like $\left| t \right|^{\frac{1}{2}}$ for a mean field $XY$-like transition. For a tricritical transition it scales like $\left| t \right|^{\frac{1}{4}}$ and is thus more rapid. For a 1st order transition there is a jump in the tilt angle upon entry to the $C$ phase.}
\label{Theta_vs_T}
\end{figure}

Figure \ref{PhaseDiagram_T_conc} shows the phase diagram for our model near the tricritical point in temperature ($T$) - concentration ($c$) space, along with the three different types of transitions: $XY$-like, tricritical and 1st order. In each case the transition from the $A$ phase to the $C$ phase implies a tilting of the optical axis away from the normal to the smectic layers by an angle $\theta$, as shown schematically in Fig. \ref{Theta_fig}. Our model gives the expected temperature dependence of $\theta$ for each type of transition, as summarized in Fig. \ref{Theta_vs_T}. For both the $XY$-like and tricritical transitions the growth of $\theta$ with decreasing temperature is continuous, although with different scaling for each transition. It should be noted that here, and throughout the article, exponents are calculated within mean field theory, and do not include the effects of fluctuations. For example, it is known that when fluctuation effects are included in analysis of the 3D $XY$ transition, $\theta$ scales like  $(1-\frac{T}{T_C})^\beta$, with $\beta\approx 0.35$, whereas in mean field theory $\beta = 0.5$. The use of mean field theory is justified by the fact that virtually all continuous $AC$ transitions are observed to be mean field like.

\begin{figure}
\includegraphics[scale=1]{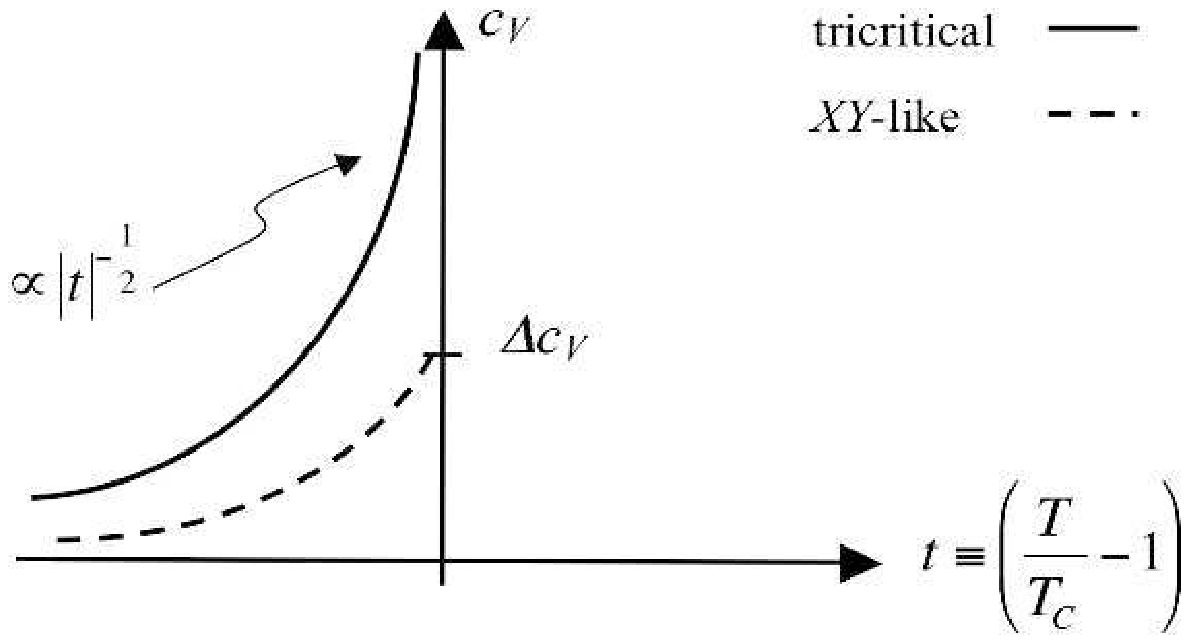}
\caption{The specific heat $c_{_{V}}$ as a function of reduced temperature $t \equiv \left(1-\frac{T}{T_C}\right)$  near the continuous $AC$ transition temperature $T_C$, i.e., for $t \ll1$. As the transition is approached from $C$ phase, the specific heat grows like $c_{_{V}} \propto \left(1-\frac{T}{T_m}\right)^{-\frac{1}{2}}$ , where $T_m>T_C$. This growth is cut off at $T=T_C$, where it reaches a maximum value, $\Delta c_{_{V}}$. If the transition becomes tricritical $T_m \rightarrow T_C$ and $c_{_{V}}$ diverges at the transition. Note that the specific heat shown here only includes the contribution from the piece of the free energy density associated with the ordering as the system moves into the $C$ phase. For a 1st order transition there will be a latent heat absorbed in going from the $C$ phase to the $A$ phase.}
\label{Specific_Heat_Intro}
\end{figure}

For the 1st order transition the tilt angle $\theta$ jumps discontinuously at the transition. Our model also leads to the expected \cite{Huang&Viner} temperature dependence of specific heat $c_{_V}$ near the continuous $AC$ transition. This temperature dependence is shown in Fig. \ref{Specific_Heat_Intro}. For an $XY$-like transition $c_{_V}$ jumps by an amount $\Delta c_{_V}$ as the system enters the $C$ phase. If the transition becomes tricritical ($M_0 \rightarrow M_{{TC}+}$, via decreasing concentration), the size of this jump diverges. Our model predicts that the divergence should scale like
\begin{eqnarray}
\Delta c_{_V} \propto \frac{1}{M_0 -M_{TC}} \;.
\label{SpecificHeatJump}
\end{eqnarray}
For a 1st order $AC$ transition there is an associated latent heat $l$. We show that if the transition becomes tricritical ($M_0 \rightarrow M_{{TC}-}$, via increasing concentration) then the latent heat vanishes like 
\begin{eqnarray}
l\propto (M_{TC}-M_0)\;.
\label{LatentHeatVanish}
\end{eqnarray}

\begin{figure}
\includegraphics[scale=1]{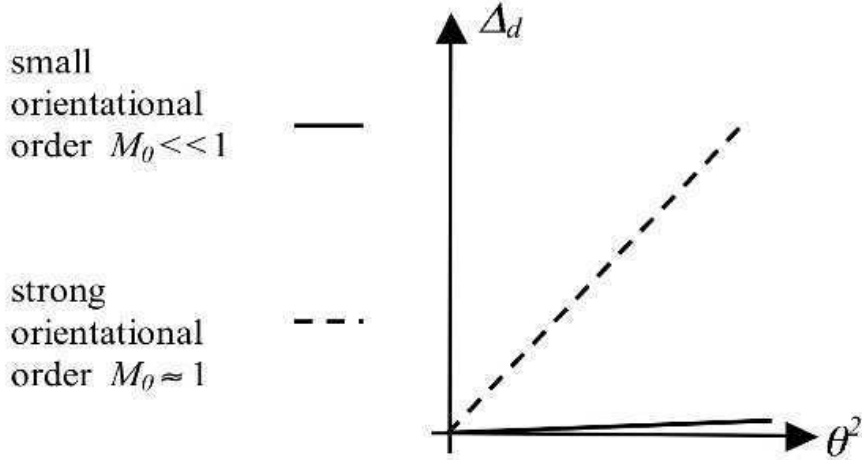}
\caption{The layer contraction $\Delta_d \equiv (d_{AC}-d_C)/d_{AC}$ as a function of $\theta^2$ near the $AC$ transition. For any type of transition the contraction will scale like $M_0 \theta^2$. Thus, the slope of $\Delta_d$ versus $\theta^2$ is proportional to the orientational order $M_0$ in the system. Near tricriticality, the orientational order is small and $M_0 \ll1$ and so the contraction is also small. Also shown is the layer contraction for a system with strong orientational order $M_0 \approx 1$, for which the contraction will be sizable. For a 1st order transition there will be a jump in the tilt angle $\theta$ at the transition and thus, the $\Delta_d$ vs.  $\theta^2$ line does not extend all the way to zero. 
}
\label{Delta_d_fig}
\end{figure}

The model is also used to examine the behavior of the layer spacing and birefringence for the three possible transitions ($XY$-like, tricritical, 1st order). We show that, for all three types of transitions, an unusually small layer contraction can be directly attributed to unusually small orientational order, $M_0$. Specifically, we find that for any of the three possible types of transitions
\begin{eqnarray}
\Delta_d \propto M_0 \left(1-\cos(\theta)\right)\approx \frac{1}{2} M_0 \theta^2\;,
\label{Contraction}
\end{eqnarray}
where the tilt angle $\theta$ is small near a continuous or weakly 1st order transition. We define the layer contraction as $\Delta_d \equiv (d_{AC}-d_C)/d_{AC}$, where $d_{AC}$ and $d_C$ are the values of the layer spacing in the $A$ phase (right at the $AC$ boundary) and in the $C$ phase, respectively. Schematic plots of  $\Delta_d$ vs. $\theta^2$ are shown in Fig. \ref{Delta_d_fig} for two types of systems: one `` de Vries"-like and the other ``conventional" . The ``de Vries"-like system has small orientational order $M_0 \ll 1$ and thus has a small slope of $\Delta_d$ vs. $\theta^2$, which corresponds to small layer contraction. The ``conventional" system has strong orientational order $M_0 = O(1)$, and thus has a larger slope, which corresponds to significant layer contraction. It should be noted that for a 1st order transition there will be a jump in the tilt angle $\theta$ at the transition, and thus, the $\Delta_d$ versus  $\theta^2$ line would not extend all the way to zero. 

\begin{figure}
\includegraphics[scale=0.9]{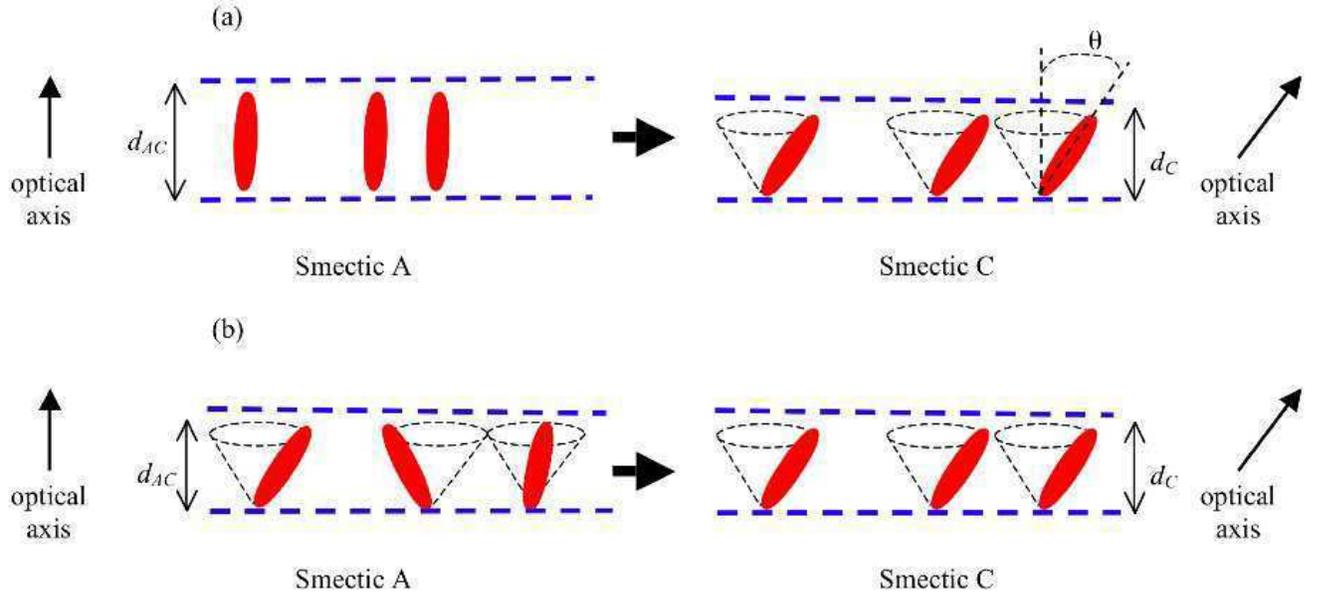}
\caption{(a) An oversimplified schematic showing the arrangement of molecules in the $A$ phase, in which the orientational order is perfect. Such a model predicts that, as the system moves into the $C$ phase, the layer spacing should contract according to $\Delta_d \equiv \left(1-\cos(\theta) \right)$, where $\Delta_d = (d_{AC}-d_C)/d_{AC}$. (b) A more realistic arrangement of the molecules in which the molecular axes are tilted away from the optical axis, but in azimuthally random directions. The more that the molecules are tilted, the smaller the orientational order. As the system moves into the $C$ phase, the ``pre-tilted" molecules do not need to tilt but rather need only to order azimuthally, thus leading to an unusually small layer contraction. Thus, the smaller the orientational order in the $A$ phase, the more ``pre-tilted" the molecules will be and the smaller the layer contraction will be, an interpretation consistent with our result, Eq. (\ref{Contraction}). The figure also shows that, as a result of the azimuthal ordering as the system moves into the $C$ phase, it should become more orientationally ordered.}
\label{Geometric Arguement}
\end{figure}

This result of our rigorous theory complements the simple geometric diffuse cone argument of de Vries \cite{DiffuseCone}, which is shown in Fig. \ref{Geometric Arguement}. The conventional, but oversimplified, relationship between layer contraction and tilt angle, $\Delta_d = \left(1-\cos(\theta)\right)$, is obtained geometrically by assuming a liquid crystal with perfect orientational order, as shown in Fig. \ref{Geometric Arguement}(a). However, it has long been known that the orientational order in liquid crystals is far from perfect. The schematic in Fig. \ref{Geometric Arguement}(b) shows a more realistic arrangement of the molecules in the $A$ phase. The molecular axes are tilted away from the optical axis, but in azimuthally random directions. One can see that the more the molecules are tilted, the smaller the orientational order in the $A$ phase. The diffuse cone model argues that, upon entry to the $C$ phase, the ``pre-tilted'' molecules do not need to tilt but rather need only to order azimuthally, thus leading to an unusually small layer contraction. Thus, the smaller the orientational order in the $A$ phase, the more ``pre-tilted'' the molecules will be and the smaller the layer contraction will be. As shown in Eq. (\ref{Contraction}), our rigorous theoretical analysis predicts a small contraction for systems with small orientational order, which agrees with this geometric argument. It also correlates well with the general experimental observation \cite{de Vries review} that de Vries materials have small orientational order. 

\begin{figure}
\includegraphics[scale=1]{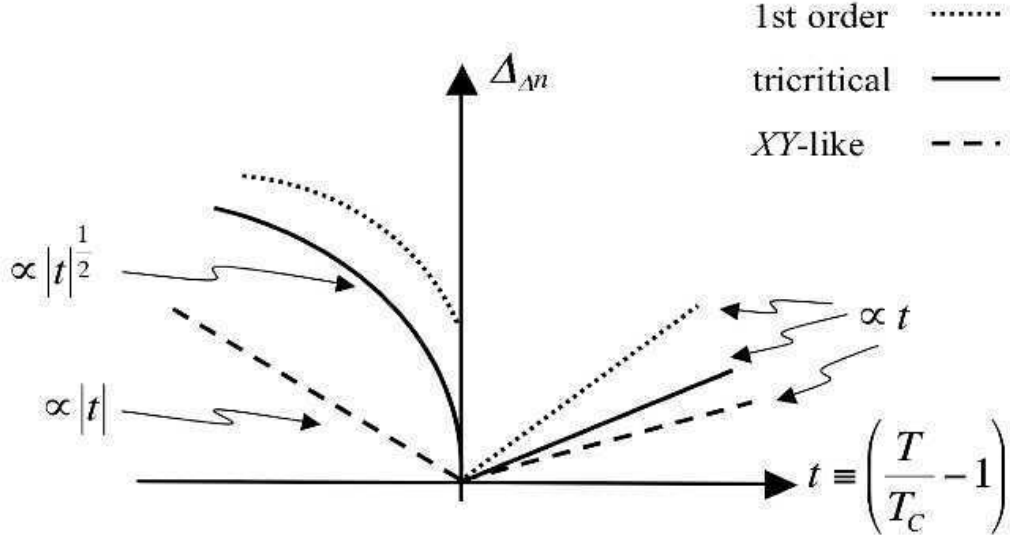}
\caption{The fractional change in birefringence $\Delta_{\Delta n} \equiv \frac{\Delta n -\Delta n_{AC}}{\Delta n_{AC}} $ as a function of reduced temperature $t \equiv \left(1-\frac{T}{T_C}\right)$ near the $AC$ transition temperature $T_C$, i.e., for $t \ll1$. For materials with excluded volume interactions, we expect the birefringence $\Delta n$, and thus $\Delta_{\Delta n}$, to {\it decrease} as the $AC$ transition is approached from within the $A$ phase. For all three types of transitions ($XY$-like, tricritical, 1st order) this decrease will scale linearly $\propto t $ with reduced temperature. Upon entry to the $C$ phase the birefringence  $\Delta n$, and thus $\Delta_{\Delta n}$, will grow. The growth is linear $\propto \left| t \right|$ for a mean field $XY$-like transition. For a tricritical transition the growth scales like $\propto \left| t \right|^{\frac{1}{2}}$ and is thus more rapid. For a 1st order transition there will be a jump in birefringence as the system enters the $C$ phase. 
}
\label{Birefringence_vs_T}
\end{figure}

From Fig. \ref{Geometric Arguement}(b) one also expects a growth of orientational order, and hence birefringence $\Delta n$, as the system moves into the $C$ phase. It is useful to define a fractional change in birefringence $\Delta_{\Delta n} \equiv \frac{\Delta n -\Delta n_{AC}}{\Delta n_{AC}} $, where $\Delta n _{AC}$ is the value of the birefringence in the $A$ phase right at the $AC$ boundary. Our model predicts that upon entry to the $C$ phase, for any of the three types of transitions ($XY$-like, tricritical, 1st order), $\Delta_{\Delta n} $ of a de Vries type material will grow according to $\Delta_{\Delta n} \propto \theta^2$. While the dependence of $\Delta_{\Delta n}$ on $\theta$ is the same for all three types of transitions, its dependence on temperature is not the same because, as shown in Fig. \ref{Theta_vs_T},  $\theta$ scales differently with temperature for each type of transition. Thus,
\begin{eqnarray}
\Delta_{\Delta n} \propto \theta^2 \propto
\begin{cases}
 (1-\frac{T}{T_C})& \text{$XY$-like} \\
 (1-\frac{T}{T_C})^{\frac{1}{2}} & \text{tricritical}\\
\text{jump} & \text{1st order}
\end{cases}&\;.
\label{birefringence}
\end{eqnarray}
The growth of $\Delta_{\Delta n}$ as a function of reduced temperature $t\equiv \left(\frac{T}{T_C}-1\right)$ is shown in Fig. \ref{Birefringence_vs_T}. For an $XY$-like transition the growth will be linear $\propto \left| t \right|$, while for a transition at tricriticality it scales like $\propto \left| t \right|^{\frac{1}{2}}$ and is thus more rapid. For a 1st order transition there will be a jump in the tilt angle and thus an associated jump in $\Delta_{\Delta n}$, although near tricriticality, where the transition is only weakly 1st order, the jump will be small. 

Our model also predicts (for materials with excluded volume interactions) the possibility of birefringence that {\it decreases} as the $AC$ transition is approached from the $A$ phase, which as discussed above, is an unusual feature that has been observed experimentally \cite{Lagerwall, YuriPrivate}. For any of the three types of transitions $\Delta_{\Delta n}$ decreases linearly with temperature as the transition is approached from the $A$ phase, as shown in Fig. \ref{Birefringence_vs_T}. The decrease in birefringence is particularly unusual, as it indicates that the system is becoming less ordered (orientationally) as a lower symmetry ($C$) phase is approached. To the best of our knowledge, this is the first example of such a phenomenon.

It should be emphasized that our analysis is only made tractable, and thus is only valid, in the limit of weak coupling between order parameters. This means that our results do not imply that all materials with small orientational order will have $AC$ transitions close to tricriticality or will exhibit de Vries behavior. Similarly, not all materials exhibiting de Vries behavior must have $AC$ transitions near tricriticality. In other words, the conclusions that our model leads us to are generic but not ubiquitous.
The remainder of this article is organized as follows. In Section \ref{Model} we introduce our model and in Section \ref{Biaxiality Induced $AC$ Tricritical Point} we locate and analyze the biaxiality induced tricritical point. We then analyze the nature ($XY$-like, tricritical, 1st order) of the $AC$ transition near this tricritical point in Section \ref{$AC$ Transition Near The Tricritical Point}. In Section \ref{Thermodynamic nature of the $AC$ transition near tricriticality} we examine the thermodynamic nature of each type of transition. Specifically, we calculate the specific and latent heats for the continuous and 1st order  transitions, respectively. Lastly we study the behavior of the layer spacing and birefringence near the $AC$ transition in Section \ref{Behavior of the Layer Spacing and Birefringence near the $AC$ transition}. We briefly summarize our results in Section \ref{Summary}. The Appendix includes details of the analysis from Section \ref{Behavior of the Layer Spacing and Birefringence near the $AC$ transition}.

\section{Model}
\label{Model}

\begin{figure}
\includegraphics[scale=1]{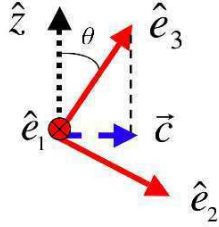}
\caption{The unit eigenvectors, ${\bf \hat e_1}$, ${\bf \hat e_2}$, ${\bf \hat e_3}$ of the orientational order tensor ${\cal Q}$. These are shown as solid arrows, with ${\bf \hat e_1}$ pointing into the page. Also shown, as a dotted arrow, is the layering direction ${\bf \hat z}$, which is normal to the plane of the layers. The eigenvector ${\bf \hat e_3}$ corresponds to the average direction of the molecules' long axes. The order parameter, ${\bf c}$, for the $C$ phase is the projection of ${\bf \hat e_3}$ onto the plane of the layers, and is shown as a dashed arrow. The angle $\theta$, by which the optical axis tilts, is also shown.}
\label{Eigenvectors}
\end{figure}

The starting point for our analysis is a generalized version of the free energy density introduced in Ref. \cite{Saunders},  which includes orientational, tilt (azimuthal), biaxial and layering order parameters. The
complex layering order parameter $\psi$ is defined via the density $\rho=\rho_0+$ Re$(\psi e^{i \bf q \cdot r})$ with $\rho_0$ constant and ${\bf q}$ the layering wavevector, the arbitrary direction of which is taken to be $z$.  The remaining order parameters are embodied in the usual second rank tensor orientational order parameter $\cal Q$, which is most conveniently expressed as
\begin{eqnarray}
Q_{ij} = M [ (-\cos(\alpha)+\sqrt{3}\sin(\alpha))e_{1i} e_{1j} \nonumber\\ 
+(-\cos(\alpha)-\sqrt{3}\sin(\alpha))e_{2i} e_{2j} \nonumber\\ 
+2\cos(\alpha)e_{3i} e_{3j}] \;,
\label{Q}
\end{eqnarray}
where ${\bf \hat e_3} = {\bf c} + \sqrt{1-c^2}{\bf \hat z}$ is the average direction of the molecules' long axes, (i.e., the director). Here, in either smectic phase, ${\bf \hat z}$ is normal to the plane of the layers.
The projection, ${\bf c}$, of the director onto the layers is the order parameter for the $C$ phase. The other two principal axes of ${\cal Q}$ are given by ${\bf \hat e_1} = {\bf \hat z} \times {\bf \hat c}$ and ${\bf \hat e_2} = \sqrt{1-c^2}{\bf \hat c} - c {\bf \hat z}$. These unit eigenvectors are shown in Fig. \ref{Eigenvectors}. The amount of orientational order is given by $M\propto \sqrt{ Tr({\cal Q}^2)}$, which is thus proportional to the birefringence. The degree of biaxiality is described by the parameter $\alpha$. The $A$ phase is untilted (${\bf c} = {\bf 0}$) and uniaxial ($\alpha=0$), while the $C$ phase is tilted (${\bf c} \neq {\bf 0}$) and biaxial ($\alpha\neq 0$). From Fig. \ref{Eigenvectors} it can be seen that the angle $\theta$, by which the optical axis tilts, can be related to $c$ via $c=\sin(\theta)$.
Taking both $\psi$ and $\cal Q$ to be spatially uniform allows the use of a Landau free energy
density $f= f_{Q}+f_{\psi}+f_{Q \psi} $, with the orientational ($f_Q$), layering ($f_\psi$), and coupling ($f_{Q \psi}$) terms given by
\begin{eqnarray}
f_{Q} =\frac{t_n Tr({\cal Q}^2)}{12} - \frac{w Tr({\cal
Q}^3)}{18} + \frac{u_n (Tr({\cal Q}^2))^2}{144} \;,
\label{H_Q}
\end{eqnarray}
\begin{eqnarray}
f_{\psi} = \frac{1}{2} t_s |\psi|^2 + \frac{1}{4} u_s |\psi|^4
+\frac{1}{2} K (q^2-q_0^2)^2|\psi|^2 ,
\label{f_Psi}
\end{eqnarray}
\begin{eqnarray}
f_{Q \psi} = \frac{q_i q_j |\psi|^2}{2} \bigg[-(a(q^2)-b(q^2)|\psi|^2)Q_{ij} + g(q^2) Q_{ik} Q_{jk} \nonumber\\ + \frac{h(q^2)}{2} q_k q_ l Q_{kl} Q_{ij} - \frac{s(q^2)}{4} (q_k q_ l Q_{kl})^2 Q_{ij} 
\bigg]
\;,
\label{H_Q_Psi}
\end{eqnarray}
where the Einstein summation convention is implied and $q_i \equiv q \delta_{iz}$. As usual in Landau theory, the parameters $t_n$ and $t_s$ are monotonically increasing functions of temperature and control the ``bare'' orientational and layering order parameters, $M_0$ and $\psi_0$ respectively. By ``bare'' we mean the values the order parameters would take on in the absence of the coupling term $f_{Q \psi}$. Similarly, the constant $q_0$ is the bare value of the layering wavevector. From Eq. (\ref{f_Psi}) above, we immediately find  $|\psi_0|=\sqrt{-t_s/u_s}$. The remaining parameters in $f_Q$ and $f_\psi$ ($w$, $u_n$, $u_s$, $K$) are positive constants. 

The coupling piece of the free energy, $f_{Q \psi}$, includes the lowest order (in fields $\psi$ and $\cal Q$) terms necessary to obtain an $AC$ transition with tricriticality. The dependence on $q^2$ of each of the coupling parameters, $a$, $b$, $g$, $h$ and $s$,  takes into account all other possible terms that have the same tensorial form, but with higher powers of $q^2$, which is not an order parameter and is therefore not assumed to be small. For weak coupling, $q \approx q_0$ we can Taylor expand each coupling parameter, e.g.  $a(q^2) \approx a_0 +a_1(q^2-q_0^2)$, where $a_0 \equiv a(q_0^2)$, and $a_1\equiv \left.\frac{d a}{d (q^2)} \right|_{q^2=q_0^2}$. For all but one of the couplings it is sufficient to use the zeroth order approximation, e.g. $g(q^2)\approx g_0$. It will be seen below that $a_1$, the first order correction to $a_0$, is necessary for layer contraction at the $AC$ transition. For notational convenience, we will, for the remainder of the article, write $a(q^2)$ as $a$ with the $q^2$ dependence implied. To render the analysis tractable, the coupling parameters are all assumed to be small and are treated perturbatively throughout.

The relatively large number of parameters in $f$ is inevitable given the fact that the theory
incorporates four types of order, layer spacing and also allows for continuous, 1st order and tricritical $AC$ transitions. Additionally, it will be shown that proximity to tricriticality and the signatures of de Vries behavior can be interpreted simply in terms of the size of the orientational order.

\section{Biaxiality Induced $AC$ Tricritical Point}
\label{Biaxiality Induced $AC$ Tricritical Point}

To investigate the nature of the $AC$ transition, we expand the part of the free energy density involving orientational order, $f_Q+f_{Q \psi}$ in powers of the biaxial and tilt order parameters, $\alpha$ and $\bf{c}$. This expansion is done near the continuous  $AC$ transition temperature $T_C$ (i.e. for $ (T-T_C)/T_C \ll 1$) and to lowest order in $M$ and $\psi$.  We find $f_Q+f_{Q \psi} \approx f_M+f_{\text{coup}}$. The piece $f_M$ only involves the orientational order parameter $M$ and is given by
\begin{eqnarray}
f_{M} = \frac{1}{2} t_n M^2 - \frac{1}{3} w M^3 + \frac{1}{4} u_n M^4
 \;.
\label{H_M}
\end{eqnarray}
From $f_M$ we immediately find the bare value of orientational order $M_0(t_n)= (w+\sqrt{w^2-4u_n t_n})/2u_n$. It is useful to write the orientational order as a combination of the bare value and a correction: $M=M_0(1+\Delta_M)$, where the correction $\Delta_M$ is due to the coupling piece $f_{\text{coup}}$. The correction $\Delta_M$ can be thought of as an augmentation of the bare orientational order $M_0$ due to the presence of layering order. As discussed in Ref. \cite{Saunders}, de Vries behavior is implied by a virtually athermal $t_n$ (and thus, an athermal $M_0$), so that for a given material $M_0$ can be thought of as a fixed quantity. This would correspond to almost perfect excluded
volume short range repulsive molecular interactions. This means that the temperature variation in orientational order $M$ is effectively due to its coupling to the temperature dependent layering, i.e. via $\Delta_M$. We assume and verify a posteriori that in the limit of weak coupling $\Delta_M \ll 1$. Similarly, we express the wavevector as $q^2=q_0^2(1+\Delta_q)$ and the layering order as $|\psi|^2=|\psi_0|^2(1+\Delta_\psi)$. The bare wavevector $q_0$ is also taken to be athermal but the bare layering order parameter $\psi_0$ is not. 

\begin{figure}
\includegraphics[scale=1]{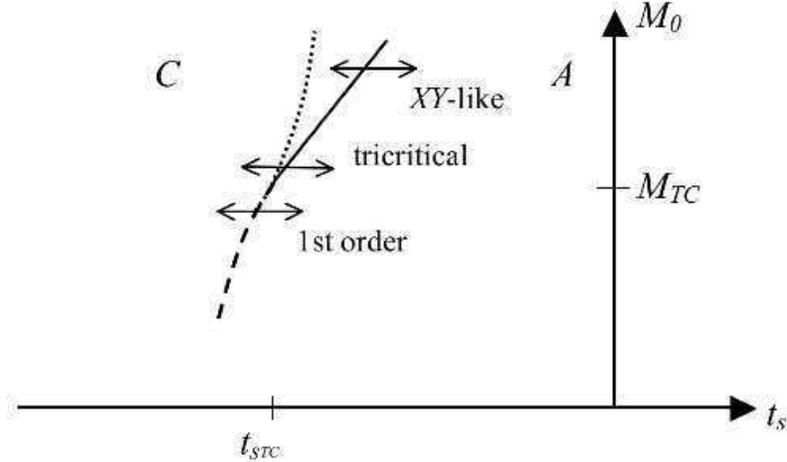}
\caption{The phase diagram in $t_s$-$M_0$ space near the tricritical point ($t_{s_{TC}}$,$M_{0_{TC}}$). The quantity $M_0$ is a measure of how much bare orientational order the system possesses and for de Vries materials is effectively athermal. Increasing concentration should increase $M_0$. The quantity $t_s$ is a monotonic function of temperature so that for a given material, decreasing the temperature corresponds to moving horizontally from right to left. The topology of the corresponding phase diagram in temperature-concentration space should essentially be the same. The solid line represents the continuous $AC$ boundary while the dashed line represents the 1st order $AC$ boundary. These two boundaries meet at the tricritical point ($t_{s_{TC}}$,$M_{0_{TC}}$). The dotted line indicates the region in which the behavior crosses over from $XY$-like to tricritical. The region in which the behavior is $XY$-like shrinks to zero as the tricritical point is approached. The slopes of the 1st order and continuous $AC$ boundaries are equal at the tricritical point. Also shown as double ended arrows, are the three distinct classes of transitions: $XY$-like, tricritical and 1st order. }
\label{Phase Diagram}
\end{figure}

The coupling piece can be broken up into three pieces: $f_{\text{coup}}=f_{M\psi} + f_c + f_{\alpha c}$. The piece $f_{M\psi}$ involves a coupling between layering and orientational order, that is non-zero in both $A$ and $C$ phases, and is given by
\begin{eqnarray}
f_{M\psi} &=& q^2  |\psi|^2 M \left(-a\tau + g_0 M - h_0 q^2 M \right)
 \;,
\label{H_M psi}
\end{eqnarray}
where
\begin{eqnarray}
\tau &=& 1-\frac{b_0 |\psi|^2 + (g_0+2h_0q^2)M}{a} \;.
\label{tau}
\end{eqnarray}
The piece $f_c$ involves the tilt (azimuthal) order parameter ${\bf c}$ and is given by
\begin{eqnarray}
f_{c} = \frac{1}{2} r_c c^2 + \frac{1}{4} u_c c^4 +\frac{1}{6} v_c c^6 \;.
\label{f_c}
\end{eqnarray}
The coefficients  $r_c$, $u_c$, $v_c$ are given by
\begin{eqnarray}
r_c &=&3 a q^2 |\psi|^2 M \tau
 \;,
\label{r_c}
\end{eqnarray}
\begin{eqnarray}
u_c &=& 9 h_0 q^4 |\psi|^2 M^2 \; ,
\label{u_c}
\end{eqnarray}
\begin{eqnarray}
v_c &=& \frac{81}{4} s_0 q^6 |\psi|^2 M^3 \;.
\label{v_c}
\end{eqnarray}
At the continuous $AC$ transition the parameter $\tau$ (and thus also $r_c$), changes sign. Close to the transition $\tau\propto(T-T_C)/T_C \ll 1$ and can be considered small. From Eq. (\ref{tau}) we see that to lowest order in the corrections $\Delta_{M, q, \psi}$ and for athermal $M_0$, this transition, occurs due to layering order increasing as temperature decreases. The transition temperature $T_C$ is defined via $|\psi_0(T_C)| = \sqrt{(a_0-(g_0+2h_0q_0^2)M_0/b_0}$, or equivalently
\begin{eqnarray}
t_s(T_C)&=&-\frac{u_s(a_0-(g_0+2h_0q_0^2)M_0)}{b_0} \;.
\label{t_s_c}
\end{eqnarray}
This continuous phase boundary is shown as a solid line in Fig. \ref{Phase Diagram}, the phase diagram  in $t_s$-$M_0$ space.  For a given material, decreasing the temperature would, in the phase diagram of Fig. \ref{Phase Diagram}, correspond to moving horizontally from right to left. The size of the orientational order $M_0$ should increase with concentration. Thus, the topology of the corresponding phase diagram, Fig. \ref{PhaseDiagram_T_conc}, in temperature-concentration space should essentially be the same as that shown in Fig. \ref{Phase Diagram}.

The coupling between tilt and biaxiality appears in the final piece
\begin{eqnarray}
f_{\alpha c} &=& A_\alpha \alpha c^2 + \frac{1}{2} B_\alpha \alpha^2 \; ,
\label{f_alpha c}
\end{eqnarray}
where, to lowest order in $\tau$,
\begin{eqnarray}
A_\alpha &=& \frac{3\sqrt{3}}{2} g_0 q^2 |\psi|^2 M^2 \; ,
\label{A}
\end{eqnarray}
\begin{eqnarray}
B_\alpha &=& 3 M^2 \left(wM-g_0 q^2 |\psi|^2 \right) \;.
\label{B}
\end{eqnarray}
From Eq. (\ref{f_alpha c}) we see that biaxiality is induced by tilt order. Minimization gives
\begin{eqnarray}
\alpha &=& - \chi_\alpha c^2\;,
\label{alpha min}
\end{eqnarray}
where $\chi_\alpha$ can be thought of as a biaxial susceptibility and is given by
\begin{eqnarray}
\chi_\alpha=\frac{\sqrt{3}}{2} \left(\frac{wM}{g_0q^2|\psi|^2}-1\right)^{-1} \;.
\label{biaxial susceptibility}
\end{eqnarray}
Keeping in mind the weak coupling regime of our analysis, i.e. $g_0 \ll 1$, we see that the systems with small orientational order $M$ will have large biaxial susceptibility. Thus, large biaxiality (and for chiral materials, an associated large spontaneous polarization) can be directly attributed to small orientational order. In fact, Eq. (\ref{biaxial susceptibility}) predicts that the biaxial susceptibility will be largest in systems that have a combination of weak orientational order ($M$) and strong layering order ($|\psi|$). It has been observed \cite{de Vries review} that this combination may be common in de Vries materials. It should be noted that the expression for $\chi_\alpha$ is only valid for $M > M_L \equiv g_0 q^2 |\psi|^2/w$, below which terms we have neglected become important. However, we will see that the tricritical point we predict occurs at a value of $M>M_L$. 

The effect of the biaxiality on the $AC$ transition is to renormalize the quartic coefficient in Eq. (\ref{f_c}), giving
\begin{eqnarray}
u_c^{\prime}= u_c\left(1-\frac{g_0}{\sqrt{3}h_0 q^2}\chi_\alpha\right)\;.
\label{renorm u_c}
\end{eqnarray}
For small biaxial susceptibility $\chi_\alpha$ (corresponding to strong orientational order), the renormalized quartic coefficient $u_c^\prime>0$ and the $AC$ transition is continuous. For large $\chi_\alpha$ (corresponding to weak orientational order), $u_c^\prime<0$ and the transition is 1st order. The tricritical point occurs at $\tau=u_c^\prime=0$, which, to lowest order in the corrections $\Delta_{q,\psi}$, corresponds to $M=M_{TC}$ with
\begin{eqnarray}
M_{TC} = \frac{a_0 g_0 q_0^2}{b_0 w} \left(1+\frac{g_0}{2h_0 q_0^2}\right)\;,
\label{M_TC}
\end{eqnarray}
which is larger than $M_L$. For small coupling ($a_0$, $b_0$, $g_0$, $h_0 \ll 1$) the value of orientational order $M_{TC}$ at tricriticality will also be small. In obtaining Eq. (\ref{M_TC}) we have used Eq. (\ref{tau}) at tricriticality to find  $|\psi_{0_{TC}}|^2 \approx a_0/b_0$, an approximation that is valid for small $M_{TC}$. Equivalently, $t_{s_{TC}} \approx -u_s a_0/b_0$.
 
\section{$AC$ Transition Near The Tricritical Point}
\label{$AC$ Transition Near The Tricritical Point}

Having found the biaxiality induced tricritical point, we now investigate the nature of the $AC$ transition in the vicinity of the tricritical point. We analyze both the continuous $AC$ transition and the 1st order $AC$ transition.

\subsection{Continous $AC$ Transition Near Tricriticality}

For sufficiently large orientational order, $M>M_{TC}$, the renormalized quartic coefficient $u_c^\prime>0$ and the $AC$ transition is continuous. As discussed in Section \ref{Biaxiality Induced $AC$ Tricritical Point}, the phase boundary is defined via $\tau=0$ or equivalently $t_s=t_s(T_C)$. Upon entry to the $C$ phase, $\tau$ becomes negative and, minimizing the effective $f_c$ (i.e. with $u_c\rightarrow u_c^\prime$) with respect to $c$ we find that the tilt order parameter grows continuously with increasing $\left| \tau \right|$ like
\begin{eqnarray}
c = \left[ \frac{2h_0^\prime}{9s_0q^2M}\left(-1+\sqrt{1+\frac{3as_0}{(h_0^\prime)^2}\left|\tau\right|}\right)\right]^{\frac{1}{2}}\;,
\label{continuous c}
\end{eqnarray}
where the effect of the coupling between biaxiality and tilt is incorporated via a renormalized $h_0^\prime$, which by expanding $\chi_\alpha$ close to tricriticality (i.e. $M \approx M_{TC}$) can be shown to be
\begin{eqnarray}
h_0^{\prime}= h_0\left(1+ \frac{2 h_0 q^2}{g_0} \right)\left(\frac{M-M_{TC}}{M_{TC}}\right)\;.
\label{renorm h_0}
\end{eqnarray}
Like $u_c^\prime$, $h_0^\prime$ changes sign at $M=M_{TC}$. It is straightforward to show that sufficiently close to the transition ($|\tau| \ll |\tau_*|$), the  dependence of $c$ on $\tau$ is effectively $XY$-like and that sufficiently far from the transition ($|\tau| \gg |\tau_*|$) it is tricritical, i.e., 
\begin{eqnarray}
c \approx
\begin{cases}
c_{_{XY}} = \sqrt{\frac{a}{3h_0^\prime q^2 M}}\left(|\tau|\right)^\frac{1}{2}& \text{$|\tau| \ll |\tau_*|$} \\
c_{_{TC}} = \left(\frac{4a}{27s_0 q^4 M^2}\right)^\frac{1}{4}\left(|\tau|\right)^\frac{1}{4} & \text{$|\tau| \gg |\tau_*|$}
\end{cases} \;.
\label{c XY and c TC}
\end{eqnarray}
The crossover from $XY$-like to tricritical behavior occurs in the region $\tau=\cal{O}(\tau_*)$ where $\tau_*$ is the value of $\tau$ where the $c_{_{XY}}=c_{_{TC}}$,
\begin{eqnarray}
|\tau_*| = \frac{4}{3}\frac{(h_0^\prime)^2}{a s_0}\;.
\label{tau C}
\end{eqnarray}
Near tricriticality where $M$ is small, the corresponding $t_{s*}$ is given by $t_{s*}=t_s(T_C)(1+|\tau_*|)$ and is shown as a dotted line in Fig. \ref{Phase Diagram}. The width of the region in which the behavior is $XY$-like shrinks to zero as the tricritical point is approached. Near the transition, the tilt angle $\theta \approx c$, and its scaling with temperature is shown in Fig. \ref{Theta_vs_T} for both an $XY$-like and a tricritical transition. Of course, the $XY$ behavior of Eq. (\ref{c XY and c TC}) is that of a mean -field theory and incorporating fluctuation effects would yield $c\propto \tau^\beta$ with $\beta\approx 0.35$. 

\subsection{1st Order $AC$ Transition Near Tricriticality}

When the orientational order is small enough ($M < M_{TC}$) the quartic coefficient ($u_c^\prime$) changes sign. The free energy now has two local minima, one at $c=0$ and another at  
\begin{eqnarray}
c_{_{1st}} = \left[ \frac{2|h_0^\prime|}{9s_0q^2M}\left(1+\sqrt{1-\frac{4\tau}{|\tau_*|}}\right)\right]^{\frac{1}{2}}\;.
\label{1st order c}
\end{eqnarray}
The 1st order $AC$ transition, and the jump from $c=0$ to $c=c_{_{1st}}$, occurs when the free energy at $c_{_{1st}}$ becomes smaller than the free energy at $c=0$. The location of the 1st order boundary can thus be obtained by finding where the two free energies are equal, or equivalently, where the difference $\Delta f$ between them is zero. To lowest order in corrections $\Delta_{M,q,\psi}$ this difference is just the effective $f_c$ (i.e. with $u_c\rightarrow u_c^\prime$) evaluated at $c_{_{1st}}$ and is given by  
\begin{eqnarray}
\Delta f = \frac{|h_0^\prime|^3}{27s_0^2}\left(1+\sqrt{1-\frac{4\tau}{|\tau_*|}}\right)^2\left(1-2\sqrt{1-\frac{4\tau}{|\tau_*|}}\right)\;, 
\label{Delta f}
\end{eqnarray}
which when set to zero yields an expression for the location of the 1st order $AC$ boundary
\begin{eqnarray}
\tau_{_{1st}}=\frac{3}{16}|\tau_*|\;.
\label{tau_1st}
\end{eqnarray}
This boundary is shown as a dashed line in Fig. \ref{Phase Diagram}. At the transition the tilt order parameter jumps from zero to a value $c_{_{1st_{AC}}}=\sqrt{|h_0^\prime|/(3s_0 q^2 M)}$. Close to tricriticality, where the transition is weakly 1st order, $c_{_{1st}}$ is small and $\approx \theta$. The corresponding temperature dependence of $\theta$ is shown in Fig. \ref{Theta_vs_T}. The size of the jump in $c$ (and thus $\theta$) goes to zero at the tricritical point, where $h_0^\prime \rightarrow 0_{-}$.

\section{Thermodynamic nature of the $AC$ transition near tricriticality}
\label{Thermodynamic nature of the $AC$ transition near tricriticality}

We next investigate the thermodynamic nature of the $AC$ transition near tricriticality. First we analyze the specific heat near the continuous transition and then the latent heat at the 1st order transition.

\subsection{Specific heat near the continuous $AC$ transition}

\begin{figure}
\includegraphics[scale=1]{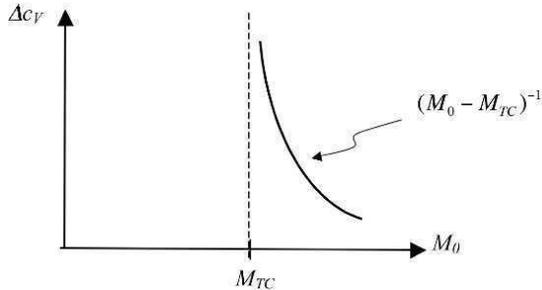}
\caption{The size of the specific heat jump $\Delta c_{_{V}}$ as a function of the system's orientational order $M_0$. As $M_0 \rightarrow M_{TC}$ the transition becomes tricritical and the specific heat jump diverges.  For systems with athermal $M_0$ it should be experimentally possible to drive the system to tricriticality by varying the concentration.} 
\label{specific heat}
\end{figure}

It is well established \cite{Huang&Viner} that the specific heat will exhibit a jump at the continuous $AC$ transition and that the thermodynamic signature of a continuous transition close to tricriticality is a divergence of this jump \cite{HuangTP}. We obtain the specific heat for our model using $c{_{_V}}=-T\frac{d^2 f_c^{\prime}}{d T^2}$, where the prime indicates the use of the biaxiality renormalized $u_c^\prime$, as given by Eq. (\ref{renorm u_c}), in $f_c$. In using $f_c^{\prime}$ instead of the full free energy density $f$, we are focussing on the contribution to the specific heat associated with the onset of ordering as the system moves into the $C$ phase. It is this contribution that is responsible for the specific heat jump. As discussed above, following Eq. (\ref{v_c}), in a material with athermal $M_0$ the transition from the $A$ to $C$ phase is driven by the layering order which increases with decreasing temperature. Near tricriticality, where the orientational order is small, the value of the layering order at the transition is $|\psi_0(T_C)|\approx\sqrt{a_0/b_0}$, and the dimensionless parameter $\tau$ can be expressed as 
\begin{eqnarray}
\tau =1-\frac{|\psi_0(T)|^2}{|\psi_0(T_C)|^2} \approx \gamma_c \left(\frac{T}{T_C}-1\right) \;,
\label{reduced temperature}
\end{eqnarray}
where we have Taylor expanded $|\psi_0(T)|$ near $T=T_C$ and the dimensionless parameter $\gamma_c>0$ is given by $\gamma_c = \left. -\frac{T_C}{|\psi_0(T_C)|^2} \frac{d |\psi_0(T)|^2}{d T} \right|_{T=T_C}$. Using Eq. (\ref{reduced temperature}), the specific heat can be expressed as 
\begin{eqnarray}
c_{_{V}} = -T\left(\frac{\gamma_c}{T_C}\right)^2\frac{d^2 f_c^{\prime}}{d \tau^2} \;.
\label{c_V definition}
\end{eqnarray}
In the $A$ phase, where $f_c^\prime=0$, the specific heat is zero. Using Eq. (\ref{continuous c}) for $c$ and Eq. (\ref{f_c}) (with $u_c \rightarrow u_c^{\prime}$) for $f_c^\prime$ we can find the specific heat in the $C$ phase. Thus we find
\begin{eqnarray}
c_{_{V}}=
\begin{cases}
0& \text{$\tau > 0$} \\
T\left(\frac{\gamma_c}{T_C}\right)^2\frac{a^2 |\psi_0(T_C)|^2}{2 h_0^{\prime}}\left[\frac{1+|\tau|}{\sqrt{1+\frac{4|\tau|}{|\tau_*|}}}+|\tau_*|\left(\sqrt{1+\frac{4|\tau|}{|\tau_*|}}-1\right)\right] & \text{$\tau < 0$}
\end{cases} \;.
\label{c_Vtau}
\end{eqnarray}
Close to tricriticality, where $\tau_*$ is small, the specific heat in the $C$ phase near the transition is dominated by the first term. Substituting $|\tau|=\gamma_c \left(1-\frac{T}{T_C}\right)$ (valid in the $C$ phase where $T<T_C$) into the first term, we find that $c_{_{V}}$ scales like
\begin{eqnarray}
c_{_{V}} \propto \left(1-\frac{T}{T_m}\right)^{-\frac{1}{2}}  \;.
\label{c_VT}
\end{eqnarray}
where $T_m=T_C\left(1+\frac{|\tau_*|}{4\gamma_c}\right)>T_C$. This scaling is shown in Fig. \ref{Specific_Heat_Intro}, where it can be seen that specific heat grows as the $AC$ transition is approached from the $C$ phase. This growth is cut off at $T=T_C$ (or equivalently $\tau=0$), where it reaches a maximum value. This maximum value is the size of the specific heat jump at the $AC$ transition and is found to be
\begin{eqnarray}
\Delta c_{_{V}} = T\left(\frac{\gamma_c}{T_C}\right)^2\frac{a^2 |\psi_0(T_C)|^2}{2 h_0^{\prime}}  \;.
\label{c_max}
\end{eqnarray}
If the transition becomes tricritical then $T_m \rightarrow T_C$ and $c_{_{V}}$ diverges at the transition. Equivalently, at tricriticality $h_0^{\prime} = 0$ and size of the jump $\Delta c_{_{V}}$ diverges. Using  Eq. (\ref{renorm h_0}) we can relate a system's bare orientational order $M_0$ to its proximity to tricriticality (where $M_0=M_{TC}$) which gives
\begin{eqnarray}
\Delta c_{_{V}} \propto \left(\frac{M_0}{M_{TC}}-1\right)^{-1} \;.
\label{c_max _M}
\end{eqnarray}
This relationship, shown in Fig. \ref{specific heat}, allows us to see how the size of the jump in specific heat would diverge if the orientational order in the system could be tuned to approach $M_{TC}$. For systems with athermal $M_0$ it should be experimentally possible to drive the system to tricriticality by varying the concentration. 

\subsection{Latent heat at the 1st order $AC$ transition}

For a 1st order $AC$ transition there will be a latent heat absorbed in going from the $C$ phase to the $A$ phase. This latent heat vanishes when the transition becomes tricritical. We obtain the latent heat $l$ for our model using $l=-T_C\frac{d f_c}{d T}$ evaluated at the 1st order boundary, where for $f_c$ we use the expression given in Eq. (\ref{Delta f}). Using the relationship between $\tau$ and $T$, as given in Eq. (\ref{reduced temperature}), we find 
\begin{eqnarray}
l= \gamma_c\frac{d f_c}{d \tau}\bigg|_{\tau=\tau_{_{1st}}} =  \gamma_c\frac{a | h_0^{\prime}|}{2 s_0} \;.
\label{latent heat}
\end{eqnarray}
As the transition becomes tricritical $h_0^{\prime} \rightarrow 0_{-}$ and the latent heat vanishes. Relating the system's bare orientational order $M_0$ to its proximity to tricriticality (where $M_0=M_{TC}$) gives
\begin{eqnarray}
l \propto \left(1-\frac{M_0}{M_{TC}}\right) \;.
\label{l _M}
\end{eqnarray}
This relationship allows us to see how the latent heat would vanish if the orientational order in the system could be tuned to approach $M_{TC}$. For systems with athermal $M_0$ it should be experimentally possible to drive the system to tricriticality, and the latent heat to zero, by varying the concentration. 

\section{Behavior of the Layer Spacing and Birefringence near the $AC$ transition}
\label{Behavior of the Layer Spacing and Birefringence near the $AC$ transition}

We next analyze the behavior of the orientational order (which is proportional to the birefringence) and the layering wavevector (which is inversely proportional to layer spacing $d$) close to the $AC$ transition. As discussed following Eq. (\ref{H_M}) above, for athermal $M_0$ and $q_0$, the temperature variation of $M=M_0(1+\Delta_M)$ and $q^2=q_0^2(1+\Delta_q)$ comes from the corrections $\Delta_M$ and $\Delta_q$ respectively. We thus seek the temperature dependence of the corrections $\Delta_{M,q}$ near the $AC$ transition. Assuming, and verifying a posteriori, that the corrections are small, we Taylor expand the free energy to order $(\Delta_{M,q})^2$ and minimize with respect to $\Delta_{M,q}$, keeping only terms to lowest order in coupling coefficients. This is done both within the $A$ phase and within the $C$ phase. Details of the analysis are given in the Appendix  \ref{Appendix}.

\subsection{Orientational order near the $AC$ transition}

For the orientational order correction within the $A$ phase we find
\begin{eqnarray}
\Delta_{M_A}=|\Delta_{M}^0|\left( - 1 + \frac{a_0}{3 g_0 M_0}\tau_0 \right)\;,
\label{Delta M_A}
\end{eqnarray}
where $\tau_0$ is just the bare value of $\tau$, i.e., $\tau$ evaluated at $M=M_0$, $\psi=\psi_0$ and $q=q_0$.  To zeroth order in corrections $\Delta_{M,\psi,q}$, $\tau=\tau_0$. The quantity $\Delta_M^0=-3 g_0 q_0^2|\psi_0(T_C)|^2/\gamma_M<0$ and for a continuous transition is just the value of the correction at the continuous $AC$ boundary, i.e., where $\tau_0=0$. At the 1st order $AC$ boundary near tricriticality, at which $\tau_0 =\tau_{_{1st}} > 0$, the correction is a little bit larger than $\Delta_M^0$ \cite{1st order footnote 1}.  Lastly, $\gamma_M= \left. d^2 f_M /d M^2\right|_{M=M_0}$. 

From Eq. (\ref{Delta M_A}) we see that as the $AC$ transition is approached from the $A$ phase, i.e. as $\tau_0\rightarrow0_{+}$, the correction $\Delta_{M_A}$ will {\it decrease}. For materials with sufficiently athermal $M_0$, this means that the orientational order will decrease as the transition is approached from above. Using the fact that birefringence $\Delta n$ is proportional to orientational order $M$, the fractional change in birefringence $\Delta_{\Delta n} \equiv \frac{\Delta n -\Delta n_{AC}}{\Delta n_{AC}} $ (where the reader is reminded $\Delta n _{AC}$ is the value of the birefringence in the $A$ phase right at the $AC$ boundary) can be related to $\Delta_M$. It is straightforward to show that, to lowest order in $\Delta_M$, $\Delta_{\Delta n}\approx \Delta_M-\Delta_M^0$. Thus, in the $A$ phase $\Delta_{\Delta n}\propto \tau_0$ will decrease as the transition is approached from above, as shown in Fig. (\ref{Birefringence_vs_T}). This is a feature that has been experimentally observed in some de Vries materials \cite{Lagerwall, YuriPrivate}. We find this feature particularly interesting, as it is the first example that we know of in which the order of a phase {\it decreases} as a transition to a lower symmetry phase is approached. It should be noted that in materials with a sufficiently strongly temperature dependent $t_N$, the growth of the ``bare'' (i.e., coupling-free) orientational order $M_0(t_n)$ as $T$ is
lowered swamps the effects due to the correction term $\Delta_{M_A}$. In this case, the orientational order would grow as the transition is approached from above. 

To find the correction near the transition within the $C$ phase one must separately analyze the three distinct regions of the phase diagram, corresponding to $XY$, tricritical and 1st order behavior. As one might expect, the dependence of $\Delta_M$ on $\tau_0\propto(T-T_C)/T_C \ll 1$ is different in each region. However, near tricriticality the dependence on the tilt order parameter $c$ in each respective region (i.e. $c_{_{XY}}$, $c_{_{TC}}$ and $c_{_{1st}}$) is identical and is given by
\begin{eqnarray}
\Delta_{M_C}=|\Delta_M^0|\left( - 1 + \frac{1}{2}\left(1+\frac{2h_0q_0^2}{g_0}\right)c^2 \right) \;,
\label{Delta_M_C}
\end{eqnarray}
where $\Delta_M^0$ is equal to the value of the correction in the $A$ phase right at the transition \cite{1st order footnote 1}. In each of the three regions the orientational order grows as one moves into the $C$ phase, consistent with birefringence measurements of de Vries materials. Using the fact that the optical axis tilt angle $\theta \approx c$ near the transition, we predict that the fractional change in birefringence will grow like $\Delta_{\Delta n}\propto \theta^2$.  It is important to note that while the dependence of the growth of $\Delta_{\Delta n}$ on $\theta$ is the same in each of the three distinct regions of the phase diagram, the dependence on $\tau_0$ is not. This is because the dependence of $c$ (and thus $\theta$) on $\tau_0$ differs in each of the three regions. For sufficiently large orientational order, away from the tricritical point $c\propto |\tau_0|^{\frac{1}{2}}$ and the growth of $\Delta_{\Delta n}$ near the continuous transition will scale like $(T_C-T)$. For smaller orientational order, near the tricritical point $c\propto |\tau_0|^{\frac{1}{4}}$ and the growth of $\Delta_{\Delta n}$ will scale like $(T_C-T)^{\frac{1}{2}}$. These scalings are shown in Fig. \ref{Birefringence_vs_T}. Thus, our model predicts that for continuous transitions near tricriticality one will see a particularly rapid growth of birefringence as one moves into the $C$ phase. For a 1st order transition there will be a jump in $c$ and thus an associated jump in the birefringence. Close to the tricritical point, where the transition is weakly 1st order, this jump will be small. 

\subsection{Layer spacing near the $AC$ transition}

For the layering wavevector (which is inversely proportional to the layer spacing) within the $A$ phase we find that 
\begin{eqnarray}
\Delta_{q_A}=\Delta_q^0 + \frac{a_0 M_0}{K q_0^2}\tau_0\;,
\label{Delta q_A}
\end{eqnarray}
where $\Delta_q^0= a_1 M_0/ K$ is value of the correction at the continuous $AC$ boundary and the reader is reminded that $a_1 = \left. \frac{d a}{d (q^2)} \right|_{_{q^2=q_0^2}}$. At the 1st $AC$ boundary near tricriticality, at which $\tau_0 =\tau_{_{1st}} > 0$, the correction is a little bit larger than $\Delta_q^0$ \cite{1st order footnote 2}. From the above equation we see that as the $AC$ transition is approached, i.e. as $\tau_0\rightarrow0_{+}$, the layering wavevector decreases. This corresponds to the layer spacing increasing, a feature which is generally observed experimentally.

As with the orientational order, it is necessary to separately analyze the three distinct regions ($XY$, tricritical and 1st order) of the phase diagram to obtain the correction near the $AC$ boundary in the $C$ phase. Similarly, while the dependence of this correction on $\tau_0$ differs within each region, the dependence on the respective tilt order parameter $c$ in each region (i.e. $c_{_{XY}}$, $c_{_{TC}}$ and $c_{_{1st}}$) is identical. It is given by
\begin{eqnarray}
\Delta_{q_C}=\Delta_q^0+\frac{3|a_1| M_0}{2K}c^2\;,
\label{Delta_q_C}
\end{eqnarray}
where $\Delta_q^0$ is equal to the value of the correction in the $A$ phase right at the transition \cite{1st order footnote 2} and for a layer contraction (as opposed to dilation) to occur we have required $a_1<0$. Using the above equation and the relationship between layer spacing ($d$) and wavevector ($q=2\pi/d$) we next seek the contraction in the layer spacing. This contraction is defined as  $\Delta_d = (d_{AC}-d_C)/d_{AC}$, where $d_{AC}$ and $d_C$ are the values of the layer spacing in the $A$ phase (right at the $AC$ boundary) and in the $C$ phase respectively. We find that this contraction is given by
\begin{eqnarray}
\Delta_d=\frac{3|a_1| M_0}{2K}c^2\;.
\label{Delta_d}
\end{eqnarray}
Near the transition $\theta \approx c$ and the fractional contraction scales like $\theta^2$, as one would expect from the simple geometric argument discussed in the Introduction. However, our theory predicts that this fractional contraction is also proportional to the size of the orientational order, $M \approx M_0$. Thus, systems with unusually small orientational order will exhibit an unusually small layer contraction, as shown in Fig. \ref{Delta_d_fig}. Given the fact that the tricritical point predicted by our model also occurs for small orientational order, it would not be surprising for some de Vries materials to exhibit $AC$ transitions close to tricriticality. It should also be noted that for the 1st order transition, the contraction will be discontinuous, although the size of the discontinuity will nonetheless be proportional to the orientational order, which if small will make the contraction small.

\section{Summary}
\label{Summary}

In summary, we have shown that our generalized Landau theory exhibits a biaxiality induced $AC$ tricritical point. The effect of the biaxiality is larger in systems with small orientational order, which would correspond to systems with narrow $A$ phases. This means that the two mechanisms that have been proposed as leading to tricriticality in a system, the coupling of tilt to biaxiality and the width of the $A$ phase, can both be attributed to the system possessing sufficiently small orientational order. For materials with excluded volume interactions, one could reduce the orientational order, and thus access a tricritical point, by reducing concentration. We have shown that the optical tilt, specific heat and latent  heat all exhibit the expected behavior near tricriticality. In addition, we have explored the effect of proximity to tricriticality on these quantities, and we have quantified the effect in terms the degree of orientational order in the system. 

We have also analyzed the behavior of the birefringence (via the orientational order) and the layer spacing (via the wavevector) for each of the three possible types of transitions ($XY$-like, tricritical and 1st order) near tricriticality. For de Vries material the birefringence has been shown to increase upon entry to the $C$ phase and for a continuous transition this increase is more rapid the closer the transition is to tricriticality. It was also shown that for materials with excluded volume interactions, birefringence will decrease as the $AC$ transition is approached from the $A$ phase, implying a non-monotonic temperature dependence of birefringence, a very unusual feature. We have 
used our model to obtain a relationship between the layer contraction and the tilt of the optical axis as a system moves into the $C$ phase, for any of the three possible types of transitions. This relationship predicts that systems with small orientational order in the $A$ phase will exhibit a corresponding small layer contraction. Our result correlates well with the diffuse cone geometric argument of de Vries.

Our future work in this area will involve further generalizing our model to include chirality. Having done so, we will analyze the electroclinic effect in materials near the $AC$ transition. Of particular interest will be how the size of electro-optical response depends on orientational order and proximity to a tricritical point.

\begin{acknowledgments}

We thank Matthew Moelter for a careful reading of the manuscript. Support was provided by Research Corporation in the form of a Cottrell College Science Award.

\end{acknowledgments}
 
\appendix
\section{Corrections to the Bare Orientational Order and to the Bare Layering Wavevector}
\label{Appendix}
 
In this Appendix we outline the procedure by which we obtain the corrections, $\Delta_{M}$ and $\Delta_{q}$, to the bare orientational order and to the bare layering wavevector, respectively. This is done near the $AC$ boundary for both the $A$ phase and the $C$ phase. Near the $AC$ boundary within the $C$ phase, we analyze separately the three regions of interest ($XY$-like, tricritical and 1st order).
 
\subsection{Correction to the bare orientational order}

In this section we find the correction $\Delta_{M}$  to the bare orientational order $M_0$, where $\Delta_M$ is defined via the full orientational order $M=M_0(1+\Delta_M)$. This is done by expanding the free energy to order $(\Delta_M)^2$ in the phase of interest and then finding the $\Delta_M$ that minimizes the free energy. 

\subsubsection{Correction in the $A$ phase}

We begin our analysis of the correction in the $A$ phase by expanding $f_M$, given by Eq. (\ref{H_M}),
\begin{eqnarray}
f_M\approx f_M(M_0)+\frac{1}{2}\gamma_M M_0^2(\Delta_M)^2\;,
\label{f_M_m}
\end{eqnarray}
where $\gamma_M= \left.d^2 f_M /d M^2\right|_{M=M_0}$.

In both the $A$ and $C$ phases, a non-zero $\Delta_M$ is due to the coupling parts of the free energy. In the $A$ phase only the piece $f_{M\psi}$, given by Eq. (\ref{H_M psi}), is non-zero. Expanding $f_{M\psi}$, which requires the expansion of $\tau$, yields
\begin{eqnarray}
f_{M\psi}\approx f_{{M\psi}_0}+q_0^2|\psi_0|^2M_0\left(3g_0M_0-a_0\tau_0\right)\Delta_M \;,
\label{f_M/psi_m}
\end{eqnarray}
where $f_{{M\psi}_0}$ and $\tau_0$ are the bare values of $f_{M\psi}$ and $\tau$,  i.e. evaluated at $M=M_0$, $\psi = \psi_0$ and $q=q_0$. We have ignored order $(\Delta_M)^2$ terms, which are higher order in the coupling than the $(\Delta_M)^2$ term in Eq. (\ref{f_M_m}) and are thus subdominant. Minimizing $f_M+f_{M\psi}$ with respect to $\Delta_M$ gives
\begin{eqnarray}
\Delta_{M_A}=\frac{q_0^2|\psi_0(T_C)|^2}{M_0 \gamma_M}\left(-3g_0M_0+a_0\tau_0\right)  \;,
\label{Delta_M_A-App}
\end{eqnarray}
where we have replaced $\psi_0\approx\psi_0(T_C)$ near the $AC$ transition. The above expression can be rearranged to give Eq. (\ref{Delta M_A}). From the above expression we see that the correction $\Delta_M$ is on the order of the coupling parameters, $a_0$ and $g_0$, and is thus small as was assumed in expanding the free energy.

\subsubsection{Correction in the $C$ phase}

In finding the corrections in the $C$ phase near the $AC$ boundary we first follow the same procedure as for the $A$ phase, namely the expansion of $f_M$ and $f_{M\psi}$ as given by Eqs. (\ref{f_M_m}) and (\ref{f_M/psi_m}) above. We must also expand the piece of coupling, $f_c^\prime$, that is non-zero in the $C$ phase.  The prime indicates the use of the biaxiality renormalized $u_c^\prime$, as given by Eq. (\ref{renorm u_c}), in $f_c$, which is given by Eq. (\ref{f_c}). For each separate region of interest ($XY$, tricritical and 1st order) we use the appropriate expression for $c$ in $f_c^\prime$. 

In the $XY$-like region we find
\begin{eqnarray}
f_{c_{XY}}^\prime=-\frac{r_c^2}{4u_c^\prime} = -\frac{|\psi|^2a^2\tau^2}{4h_0^\prime} \;.
\label{f_c_XY}
\end{eqnarray}
Expanding $\tau$ and $h_0^\prime$ in powers of $\Delta_M$, keeping terms to lowest order in $\tau_0$ and coupling coefficients gives
\begin{eqnarray}
f_{c_{XY}}^\prime\approx f_{c_{XY_{0}}}^\prime+\frac{|\psi_0(T_C)|^2M_0a_0\tau_0}{2h_{00}^\prime} \left(g_0+2h_0q_0^2\right)\Delta_M \;,
\label{f_c_XY_exp}
\end{eqnarray}
where $ f_{c_{XY_{0}}}^\prime$ and $h_{00}^\prime$ are the bare values of $f_{c_{XY}}^\prime$ and $h_{0}^\prime$.

Minimizing $f_M+f_{M\psi}+f_{c_{XY}}^\prime$ with respect to $\Delta_M$ gives 
\begin{eqnarray}
\Delta_{M_{C_{XY}}}=\frac{q_0^2|\psi_0(T_C)|^2}{M_0 \gamma_M}\left(-3g_0M_0
+\frac{a_0|\tau_0|}{2h_{00}^\prime q_0^2}\left(g_0+2h_0q_0^2 \right) \right)  \;,
\label{Delta_M_Cxy-App}
\end{eqnarray}
where, in neglecting the $\tau_0$ dependent contribution from $f_{M\psi}$, we have used the fact that close to tricriticality $h_{00}^\prime/h_0 \ll 1$. Using the bare version of $c=c_{_{XY}}$ as given by Eq. (\ref{c XY and c TC}) this expression can be rearranged to give Eq. (\ref{Delta_M_C}).

For the tricritical region where, $u_c^\prime$ is effectively zero, one must use $f_c^\prime$ evaluated at $c=c_{_{TC}}$ which yields
\begin{eqnarray}
f_{c_{TC}}^\prime=-\frac{1}{3}\sqrt{\frac{-r_c^3}{v_c}} = -\frac{2|\psi|^2}{3\sqrt{3}}\sqrt{\frac{-a^3\tau^3}{s_0}}  \;.
\label{f_c_TC}
\end{eqnarray}
Expanding $\tau$ in powers of $\Delta_M$ while keeping terms to lowest order in $\tau_0$ and coupling coefficients gives
\begin{eqnarray}
f_{c_{TC}}^\prime\approx f_{c_{TC_{0}}}^\prime-|\psi_0(T_C)|^2M_0\sqrt{\frac{a_0|\tau_0|}{3s_0}} \left(g_0+2h_0q_0^2\right)\Delta_M \;,
\label{f_c_TC_exp}
\end{eqnarray}
where $ f_{c_{TC_{0}}}^\prime$ is the bare value of $f_{c_{TC}}^\prime$.

Minimizing $f_M+f_{M\psi}+f_{c_{TC}}^\prime$ with respect to $\Delta_M$ gives 
\begin{eqnarray}
\Delta_{M_{C_{TC}}}=\frac{q_0^2|\psi_0(T_C)|^2}{M_0 \gamma_M}\left(-3g_0M_0+\sqrt{\frac{a_0|\tau_0|}{3s_0q_0^4}}\left(g_0+2h_0q_0^2\right)\right)  \;,
\label{Delta_M_Ctc-App}
\end{eqnarray}
where, in neglecting the $\tau_0$ dependent contribution from $f_{M\psi}$, we have used the fact that $\sqrt{\tau_0}\gg\tau_0$ close to tricriticality, i.e. where $\tau_*\ll1$. Using the bare version of $c=c_{_{TC}}$ as given by Eq. (\ref{c XY and c TC}) this expression can be rearranged to give Eq. (\ref{Delta_M_C}).

Lastly we obtain the correction in $\Delta_M$ in the $C$ phase (where $h_0^\prime<0$) near the 1st order $AC$ boundary. We do this by expanding $f_c^\prime$ near the first order $AC$ boundary, the expression for which is given by Eq. (\ref{Delta f}).  
Expanding $\tau$, $h_0^\prime$ and $\tau_*$ (which depends on $h_0^\prime$) in powers of $\Delta_M$ while keeping terms to lowest order in $\tau_0$ and coupling coefficients gives
\begin{eqnarray}
f_{c_{1st}}^\prime\approx f_{c_{1st_{0}}}^\prime-\frac{|\psi_0(T_C)|^2M_0|h_{00}^\prime|}{3s_0}\left(1+\sqrt{1-\frac{4\tau}{|\tau_*|}}\right) \left(g_0+2h_0q_0^2\right)\Delta_M \;,
\label{f_c_1st_exp}
\end{eqnarray}
where $ f_{c_{1st_{0}}}^\prime$ is the bare value of $f_{c_{1st}}^\prime$. 

Minimizing $f_M+f_{M\psi}+f_{c_{1st}}^\prime$ with respect to $\Delta_M$ gives 
\begin{eqnarray}
\Delta_{M_{C_{1st}}}=\frac{q_0^2|\psi_0(T_C)|^2}{M_0 \gamma_M}\left(-3g_0M_0+\frac{|h_{00}^\prime|}{3s_0q_0^2}\left(1+\sqrt{1-\frac{4\tau}{|\tau_{*_0}|}}\right) \left(g_0+2h_0q_0^2\right)\right)  \;,
\label{Delta_M_C1st-App}
\end{eqnarray}
where $\tau_{*_0}$ is the bare value of $\tau_*$ and, in neglecting the $\tau_0$ dependent contribution from $f_{M\psi}$, we have used the fact that close to tricriticality $h_{00}^\prime/h_0 \ll 1$. Using the bare version of $c=c_{_{1st}}$ as given by Eq. (\ref{1st order c}) this expression can be rearranged to give Eq. (\ref{Delta_M_C}).

\subsection{Correction to the bare wavevector}

In this section we find the correction $\Delta_{q}$  to the bare wavevector $q_0$, where $\Delta_q$ is defined via the full wavevector $q^2=q_0^2(1+\Delta_q)$. As with the orientational order, this is done by expanding the free energy to order $(\Delta_q)^2$ in the phase of interest and then finding the $\Delta_q$ that minimizes the free energy. 

\subsubsection{Correction in the $A$ phase}

We begin our expansion of the free energy in powers of $\Delta_q$ by expanding $f_\psi$, given by Eq. (\ref{f_Psi}),
\begin{eqnarray}
f_{\psi} \approx \frac{1}{2}K|\psi_0|^2q_0^4\Delta_q^2\;.
\label{f_psi_q}
\end{eqnarray}
In both the $A$ and $C$ phases, a non-zero $\Delta_q$ is due to the coupling parts of the free energy. In the $A$ phase only the piece $f_{M\psi}$, given by Eq. (\ref{H_M psi}), is non-zero. Expanding $f_{M\psi}$ yields
\begin{eqnarray}
f_{M\psi}\approx f_{{M\psi}_0}-q_0^2|\psi_0|^2M_0\left(a_1q_0^2+a_0\tau_0\right)\Delta_q \;,
\label{f_M/psi_q}
\end{eqnarray}
where we have used the fact that $M$ is small near tricriticality. We have ignored order $(\Delta_q)^2$ terms, which are higher order in the coupling than the $(\Delta_q)^2$ term in Eq. (\ref{f_psi_q}) and are thus subdominant. Minimizing $f_M+f_{M\psi}$ with respect to $\Delta_q$ gives
\begin{eqnarray}
\Delta_{q_A}=\frac{M_0}{Kq_0^2}\left(a_1q_0^2+a_0\tau_0\right)  \;.
\label{Delta_q_A-App}
\end{eqnarray}
The above expression can be rearranged to give Eq. (\ref{Delta q_A}). From the above expression we see that the correction $\Delta_q$ is on the order of the coupling parameters, $a_0$ and $a_1$, and is thus small as was assumed in expanding the free energy.

\subsubsection{Correction in the $C$ phase}

In finding the corrections in the $C$ phase near the $AC$ boundary we follow the same procedure as for the orientational order. To obtain the correction within the $XY$-like region we use $f_{c_{XY}}^\prime$ as given by Eq. (\ref{f_c_XY}). Expanding $\tau$ and $h_0^\prime$ in powers of $\Delta_q$, keeping terms to lowest order in $\tau_0$ and coupling coefficients gives
\begin{eqnarray}
f_{c_{XY}}^\prime\approx f_{c_{XY_{0}}}^\prime-\frac{|\psi_0(T_C)|^2a_1q_0^2a_0\tau_0}{2h_{00}^\prime}\Delta_q \;,
\label{f_c_XY_exp_q}
\end{eqnarray}
where we have used the fact that $M$ is small near tricriticality.

Minimizing $f_M+f_{M\psi}+f_{c_{XY}}^\prime$ with respect to $\Delta_q$ gives 
\begin{eqnarray}
\Delta_{q_{C_{XY}}}=\frac{a_1}{Kq_0^2}\left(M_0q_0^2-
\frac{a_0|\tau_0|}{2h_{00}^\prime}\right)  \;,
\label{Delta_q_Cxy-App}
\end{eqnarray}
where, in neglecting the $\tau_0$ dependent contribution from $f_{M\psi}$, we have used the fact that close to tricriticality $h_{00}^\prime/h_0 \ll 1$. Using the bare version of $c=c_{_{XY}}$ as given by Eq. (\ref{c XY and c TC}) this expression can be rearranged to give Eq. (\ref{Delta_q_C}).

For the tricritical region we use $f_{c_{TC}}^\prime$ as given by Eq. (\ref{f_c_TC}).  Expanding $a$ and $\tau$ in powers of $\Delta_q$ while keeping terms to lowest order in $\tau_0$ and coupling coefficients gives
\begin{eqnarray}
f_{c_{TC}}^\prime\approx f_{c_{TC_{0}}}^\prime+|\psi_0(T_C)|^2q_0^2a_1\sqrt{\frac{a_0|\tau_0|}{3s_0}} \Delta_q \;,
\label{f_c_TC_exp_q}
\end{eqnarray}
where we have used the fact that $M$ is small near tricriticality.

Minimizing $f_M+f_{M\psi}+f_{c_{TC}}^\prime$ with respect to $\Delta_q$ gives 
\begin{eqnarray}
\Delta_{q_{C_{TC}}}=\frac{a_1}{Kq_0^2}\left(M_0q_0^2-
\sqrt{\frac{a_0|\tau_0|}{3s_0}} \right)    \;,
\label{Delta_q_Ctc-App}
\end{eqnarray}
where, in neglecting the $\tau_0$ dependent contribution from $f_{M\psi}$, we have used the fact that $\sqrt{\tau_0}\gg\tau_0$ close to tricriticality, i.e. where $\tau_*\ll1$. Using the bare version of $c=c_{_{TC}}$ as given by Eq. (\ref{c XY and c TC}) this expression can be rearranged to give Eq. (\ref{Delta_q_C}).

We conclude by obtaining the correction in $\Delta_q$ in the $C$ phase (where $h_0^\prime<0$) near the 1st order $AC$ boundary. We do this by expanding $f_c^\prime$ near the first order $AC$ boundary, the expression for which is given by Eq. (\ref{Delta f}). Expanding $\tau$ and $h_0^\prime$ in powers of $\Delta_q$, keeping terms to lowest order in $\tau_0$ and coupling coefficients gives
\begin{eqnarray}
f_{c_{1st}}^\prime\approx f_{c_{1st_{0}}}^\prime+\frac{|\psi_0(T_C)|^2a_1q_0^2|h_{00}^\prime|}{3s_0}\left(1+\sqrt{1-\frac{4\tau}{|\tau_*|}}\right)\Delta_q \;,
\label{f_c_1st_exp_q}
\end{eqnarray}
where we have used the fact that $M$ is small near tricriticality. 

Minimizing $f_M+f_{M\psi}+f_{c_{1st}}^\prime$ with respect to $\Delta_q$ gives 
\begin{eqnarray}
\Delta_{M_{C_{1st}}}=\frac{a_1}{Kq_0^2}\left(M_0q_0^2-\frac{|h_{00}^\prime|}{3s_0}\left(1+\sqrt{1-\frac{4\tau}{|\tau_*|}}\right)  \right) \;,
\label{Delta_q_C1st-App}
\end{eqnarray}
where, in neglecting the $\tau_0$ dependent contribution from $f_{M\psi}$, we have used the fact that close to tricriticality $h_{00}^\prime/h_0 \ll 1$. Using the bare version of $c=c_{_{1st}}$ as given by Eq. (\ref{1st order c}) this expression can be rearranged to give Eq. (\ref{Delta_q_C}).


\begin{thebibliography}{99}
\bibitem{1st order footnote 1} It is straightforward to show that at the 1st order $AC$ transition, where $\tau=\tau_{_{1st}}$ as given by Eq. (\ref{tau_1st}), the value of the correction $\Delta_{M_{A}}=|\Delta_{M}^0|\left( - 1 + \frac{a_0}{16 g_0 M_0}\tau_{*_0} \right)$, which  
near the tricritical point is $\approx - |\Delta_M^0|$.
\bibitem{1st order footnote 2} As with the orientational order \cite{1st order footnote 1}, it is straightforward to show that at the 1st order $AC$ transition the value of the correction $\Delta_{q_{A}}=|\Delta_{q}^0|\left( - 1 + \frac{a_0}{16 |a_1| q_0^2}\tau_{*_0} \right)$, which near the tricritical point is $\approx -|\Delta_q^0|$.
\end{thebibliography}
\end{document}